\documentclass[pre,twocolumn,aps,superscriptaddress,showpacs,floatfix]{revtex4}
\usepackage{graphicx}
\usepackage{dcolumn}
\usepackage{bm}
\usepackage{amsmath}
\usepackage{amssymb}
\begin{document}
\title{Driven Brownian transport through arrays of symmetric obstacles}
\author{P. K. Ghosh}
\affiliation{Advanced Science Institute, RIKEN, Wako-shi, Saitama, 351-0198, Japan}
\affiliation{Institut f{\"u}r Physik
Universit{\"a}t Augsburg, D-86135 Augsburg, Germany}
\author{P. H\"anggi}
\affiliation{Institut f{\"u}r Physik
Universit{\"a}t Augsburg, D-86135 Augsburg, Germany}
\author{F. Marchesoni}
\affiliation{Dipartimento di Fisica, Universit\`a di Camerino,
I-62032 Camerino, Italy}
\author{S. Martens}
\affiliation{Department of Physics, Humboldt-Universit\"at zu Berlin, Newtonstr. 15, D-12489 Berlin, Germany}
\author{F. Nori}
\affiliation{Advanced Science Institute, RIKEN, Wako-shi, Saitama, 351-0198, Japan}
\affiliation{Physics Department, University of Michigan, Ann Arbor, MI 48109, USA}
\author{L. Schimansky-Geier}
\affiliation{Department of Physics, Humboldt-Universit\"at zu Berlin, Newtonstr. 15, D-12489 Berlin, Germany}
\author{G. Schmid}
\affiliation{Institut f{\"u}r Physik
Universit{\"a}t Augsburg, D-86135 Augsburg, Germany}

\begin{abstract}
We numerically investigate the transport of a suspended overdamped Brownian particle which is driven through a two-dimensional rectangular array of circular obstacles with finite radius. Two limiting cases are
considered in detail, namely, when the constant drive is parallel to
the principal or the diagonal array axes. This corresponds to studying
the Brownian transport in periodic channels with reflecting walls of
different topologies. The mobility and diffusivity of the transported particles in such channels are determined as functions of the drive and
the array geometric parameters. Prominent transport features, like
negative differential mobilities, excess diffusion peaks, and
unconventional asymptotic behaviors, are explained in terms of two
distinct lengths, the size of single obstacles (trapping length) and
the lattice constant of the array (local correlation length). Local
correlation effects are further analyzed by continuously rotating the
drive between the two limiting orientations.

\end{abstract}

\pacs{05.40.-a,05.60.Cd,51.20.+d} \maketitle

\section{Introduction}
\label{intro}

The effective control of mass and charge transport in artificial micro-
and nanostructures requires a deep understanding of the diffusive
mechanisms involving small objects in confined geometries. Such
situations are typically encountered when studying the transport of
particles in, e.g.: biological cells \cite{Hille} and zeolites
\cite{Kaerger}, catalytic reactions occurring either on templates or in
porous media \cite{Daniel}, chromatography or, more generally,
separation techniques of size-dispersed particles on micro- or even
nanoscales \cite{Corma}. In many respects these transport phenomena
can be regarded as diverse manifestations of geometrically constrained Brownian dynamics in one (1D) or higher dimensions
\cite{Brenner_book}.

In this paper we focus on the dc driven transport of pointlike
Brownian particles in two dimensional (2D) arrays. Restricting the
volume of the phase space available to the diffusing particles by
means of confining boundaries, or obstacles, causes remarkable
entropic effects
\cite{Zwanzig,Jacobs,chemphyschem,BM1,BM2,Reguera:2001,Kalinay,Laachi,Reguera:2006,Burada,gSR,Mondal}.
In quasi-1D geometries (narrow channels), driven transport of charged
particles across bottlenecks, such as ion transport through artificial nanopores or biological channels, represents an ubiquitous situation, where diffusion is effectively controlled by
entropic barriers \cite{chemphyschem}. Similarly, the operation of
artificial Brownian motors and molecular machines \cite{BM1,BM2}
also results from the interplay of thermal diffusion and the pinning
action by both energetic and entropic barriers.

\begin{figure*}[t]
\centering
\includegraphics[width=0.90\textwidth]{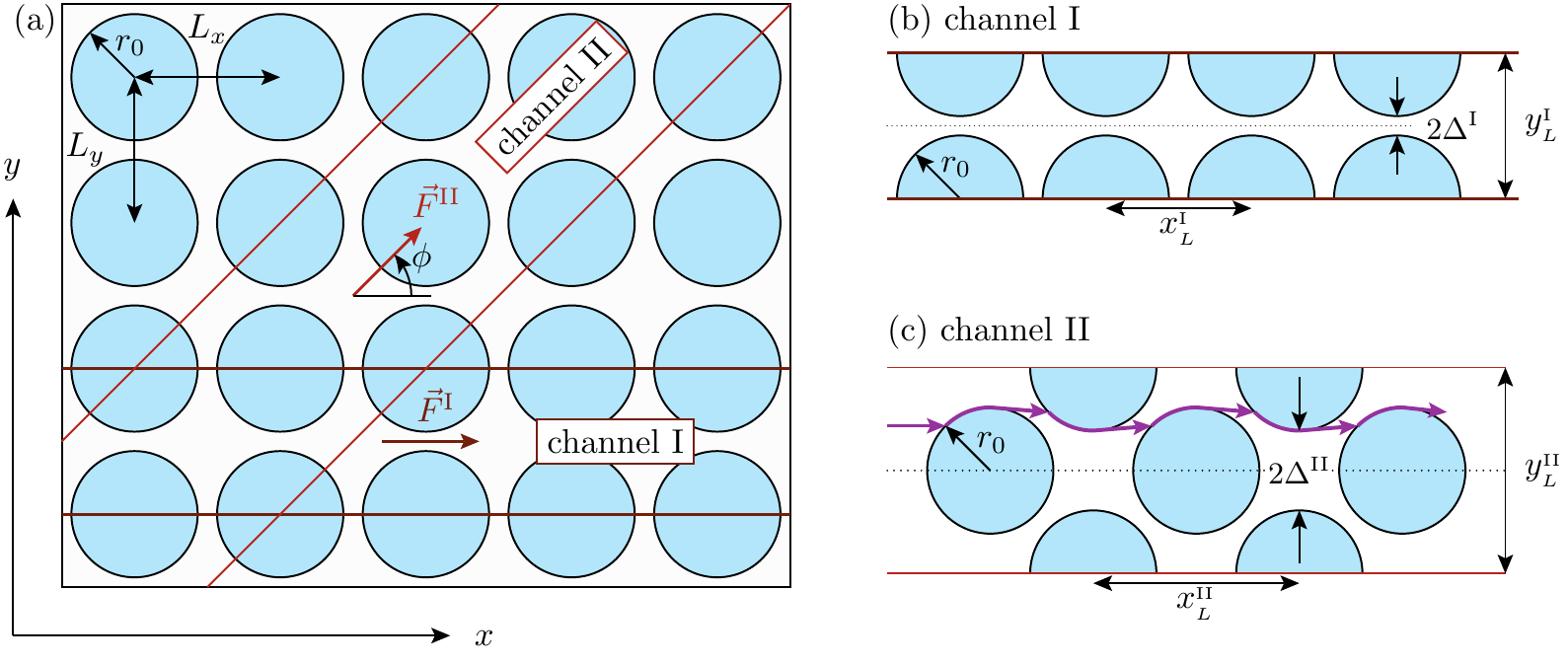}
\caption{(Color online) Sketch of a square array, $L_{x}=L_{y}$, of circular obstacles with radius $r_0$. (a) The Brownian particle is driven by a dc force $\vec {F}$ oriented at an angle $\phi$ with the horizontal axis. Transport through the array can be reduced to transport along corrugated channels of type I for $F^{\rm I}$,  $\phi=0$,  and type II for $F^{\rm II}$, $\phi=\pi/4$, with appropriate compartment sizes. Channels I and II are sketched in (b) and (c), respectively; $y^{_{\rm I, II}}_L$ and $x^{_{\rm I,II}}_L$ denote the width and the periodicity of the channel. The vertical distance between two opposite obstacles is $2\Delta^{\rm I, II} = y^{_{\rm I, II}}_L-2r_{0}$. The
trajectory drawn in (c) is the limiting noiseless trajectory for $F^{\rm II} \to \infty$ pointing to the right. The central horizontal lines in (b) and (c) are the channel symmetry axes, which
suggest to further reduce the analysis of the problem to half the sketched channels (see Sec. \ref{model}).} \label{F1}
\end{figure*}
%

Higher dimensional geometries have recently attracted broad interest in the context of separation of macromolecules \cite{Brenner}, like colloids \cite{Tierno}, DNA fragments \cite{MacDonald,Huang}, or even magnetic vortices \cite{Nori,Riken,Madrid,Moshchalkov}, because of the occurrence of induced transverse drifts, which separate different objects depending on their bulk diffusivity \cite{Drazer1,Drazer2,Drazer3}. Such systems are essentially 2D arrays of impenetrable obstacles traversed by diffusing particles subjected to external gradients. An accurate modeling of real experiments, like those cited here and many more, would require incorporating nontrivial effects due to particle interactions \cite{Nori,Drazer2}, fluidics \cite{Drazer3,Souto}, chaos \cite{Reimann}, inertia \cite{Riken}, excluded volume \cite{Brenner_book,Drazer1}, particle shape \cite{Han,Shape}, spatial asymmetries and disorder \cite{BM1,BM2}, to mention but a few. In other contexts the underlying planar constrained geometries have been modeled also by arrays of traps \cite{trap arrays,Zhu}, or egg-carton potentials \cite{Geigen,Sancho}.

In order to attempt a first quantitative characterization of forced transport across a 2D array, we used a simplified model where a single Brownian particle of negligible size is suspended in an unmovable interstitial fluid at fixed temperature (dilute suspension). The particle is free to diffuse in the connected space delimited by circular reflecting obstacles of finite radius, arranged in a rectangular lattice. The particle is subjected to thermal fluctuations and large viscous damping (as is often the case for biomolecules, colloids and magnetic vortices \cite{BM1,BM2}), and a homogeneous constant force (Fig. \ref{F1}).  Such a dc drive is applied from the outside by coupling the particle to an external field (for instance, by assuming that the particle carries a dielectric or magnetic dipole, or a magnetic flux), without inducing drag effects on the suspension fluid. To simplify our model even further, we neglected any local spatial asymmetry, such as one obtains by lining up non-circular obstacles in symmetric lattices or by arranging circular obstacles in suitable anisotropic arrays \cite{BM1,BM2}. Under such conditions the dynamics of the driven particle is fully described by an overdamped 2D Langevin equation, which we encoded in our numerical simulation algorithm, or by the corresponding 2D Fokker-Planck equation, which lends itself to a more systematic analytical treatment \cite{Brenner_book}. All other effects, including particle-particle and particle-obstacle interactions, hydrodynamic corrections, chaotic and inertial dynamical terms, particle size and shape, are thus ignored.

The properties of Brownian transport in the stationary regime are well quantified by the particle mobility and diffusivity as functions of the external drive (magnitude and orientation) and of the array
parameters (obstacle radius and lattice constants). Thanks to the simplifications assumed in the present model, we detected prominent transport features, like negative differential mobilities, excess diffusion peaks, and unconventional asymptotic behaviors, which went unnoticed in earlier reports, experimental and computational, alike. We explain the properties of Brownian transport in 2D arrays in terms of two distinct depinning mechanisms: (i) trapping by a single obstacle, which the particle can overcome only by diffusing a transverse distance of the order of the obstacle size, while being driven the same longitudinal length; (ii) correlated collisions against obstacle rows at an angle with the external force,
which tend to collimate the particle trajectories to form stream lines connecting and flowing around the obstacles. Such mechanisms
correspond to two different length scales, respectively: the obstacle radius (independent of the array geometry) and the effective obstacle spacing (resulting from the combination of array geometry and drive
orientation).

This paper is organized as follows. In Sec.~\ref{model} we introduce the Langevin equation formalism employed in our simulation code. We show how for special orientations of the drive the problem can be reduced to the transport in corrugated 1D channels. Mobility and diffusivity data are plotted in Sec.~\ref{channels} as functions of the drive, the obstacle radius, and the array lattice constants for two such corrugated channels corresponding, respectively, to orienting the drive along the principal (Sec.~\ref{chI}) and the diagonal axes of a square array (Sec.~\ref{chII}). Mobility and diffusivity dependence on the system parameters are interpreted in Sec. \ref{theory} in terms of the two aforementioned depinning mechanisms. In Sec.~\ref{angular}, local correlation effects due to the array geometry are analyzed in more detail by continuously rotating the drive between the two limiting orientations of Sec.~\ref{channels}. Finally, we summarize our results in Sec.~\ref{conclusion}.

\begin{figure}[t]
\centering
\includegraphics[width=0.41\textwidth]{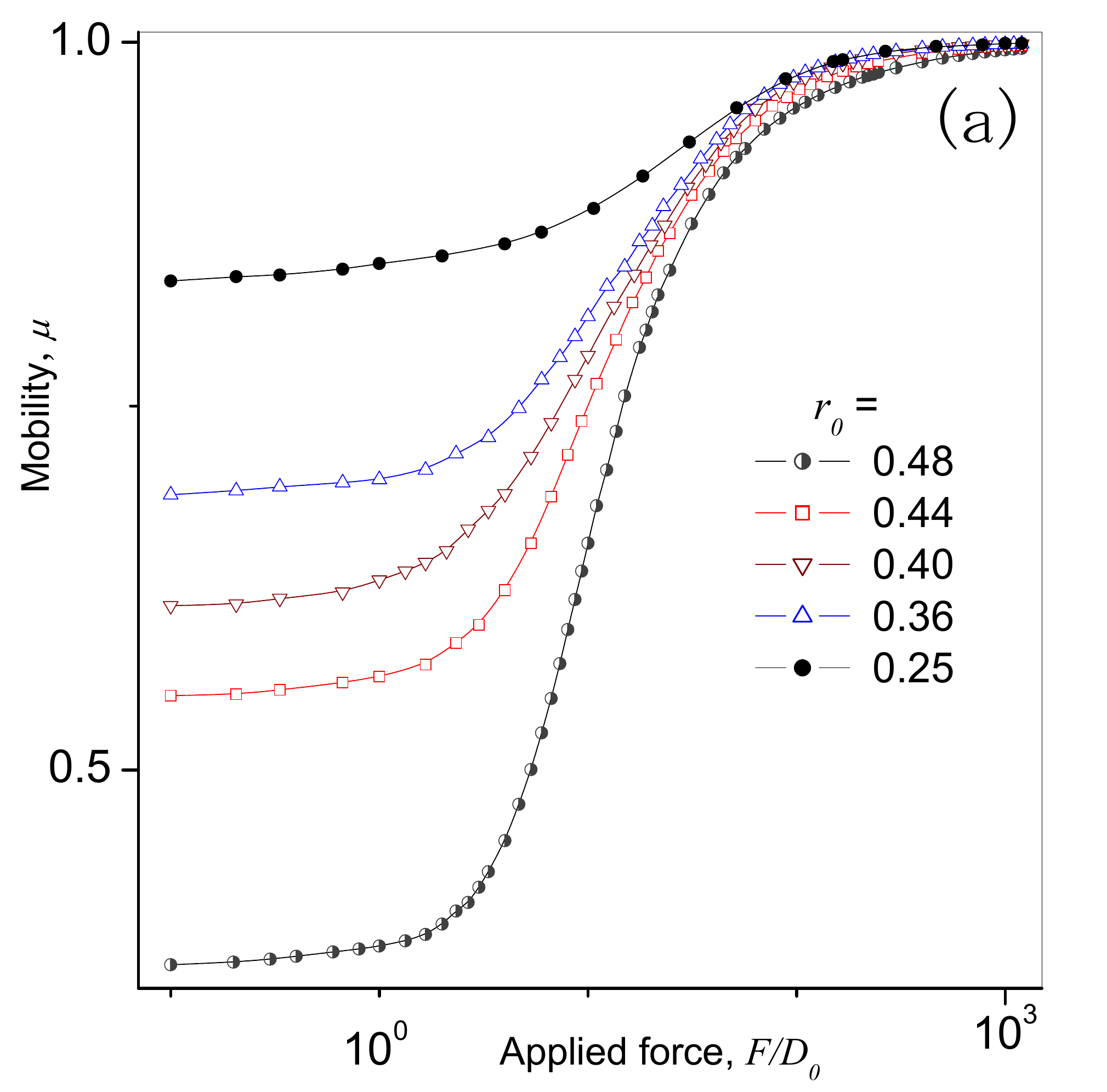}
\includegraphics[width=0.42\textwidth]{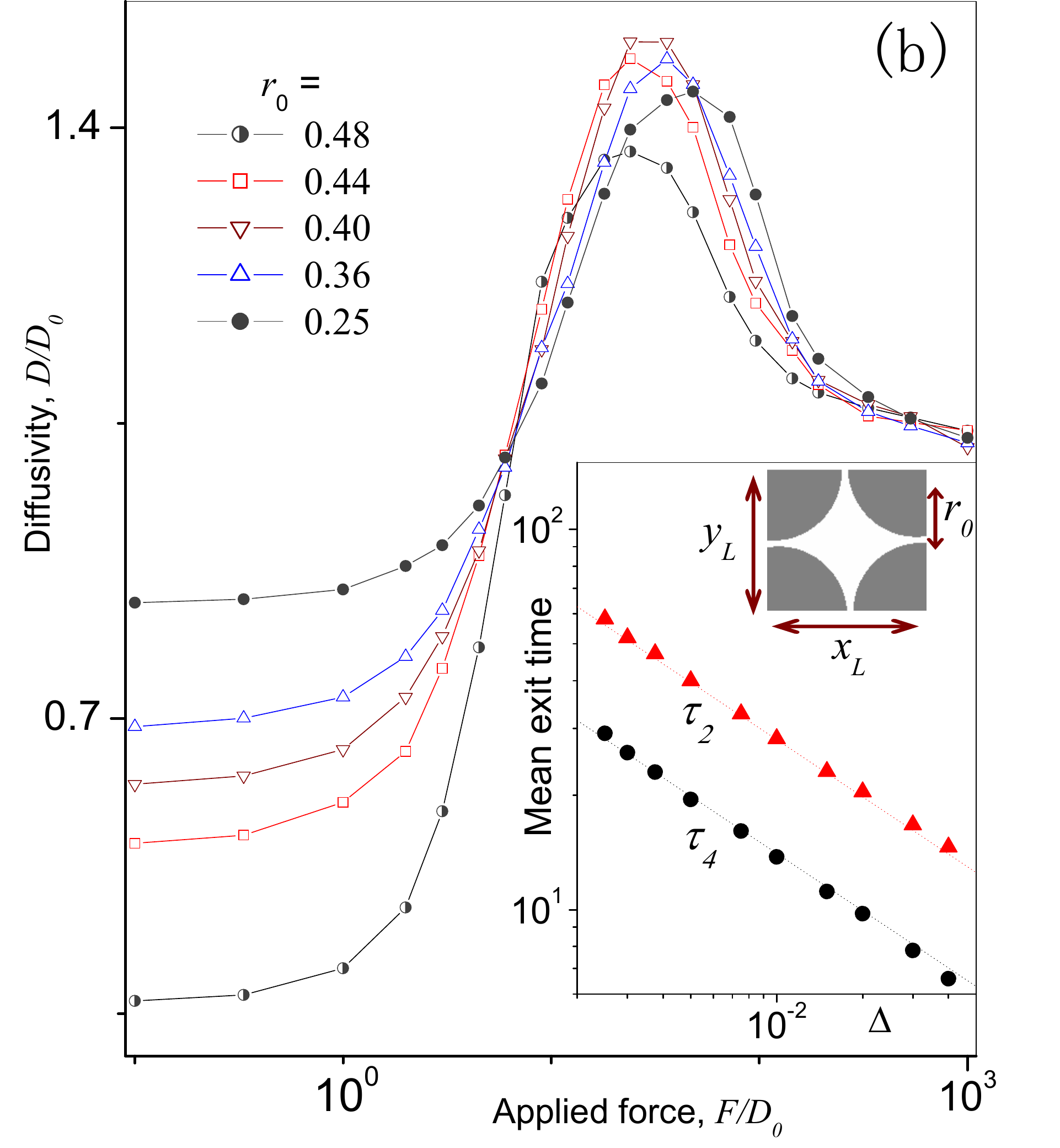}
\caption{(Color online) Driven transport in channel I with $x_L=y_L=1$, $D_0=0.03$, and different $r_0$: particle mobility $\mu$ (a) and scaled diffusivity $D/D_0$ (b) vs. the scaled force $F/D_0$. Inset: mean exit time  through four pores, $\tau_4$, and two opposite pores, $\tau_2$. The numerical data are compared with the analytical predictions in Eq. (\ref{tau2}).} \label{F2}
\end{figure}

\section{Model} \label{model}

Let us consider an overdamped Brownian particle of unit mass diffusing
in a suspension fluid contained in a 2D rectangular array, $L_{x}
\times L_{y}$, of reflecting circular obstacles of radius $r_0$, as
illustrated in Fig. \ref{F1}(a). The particle is subjected to a
homogeneous force  $\vec {F}$ oriented at an angle $\phi$ with the
horizontal axis $x$. The overdamped dynamics of the particle is
modeled by the 2D Langevin equation,
\begin{equation}\label{le}
\frac{d{\vec r}}{dt}={\vec F}\;+ \sqrt{D_0}~{\vec \xi}(t),
\end{equation}
where ${\vec r}=(x,y)$ and ${\vec \xi}(t)=(\xi_x(t),\xi_y(t))$ are zero-mean, white Gaussian noises with autocorrelation functions $\langle \xi_i(t)\xi_j(t')\rangle = 2\delta_{ij}\delta(t-t')$, with
$i,j=x,y$. The noise strength $D_0$ corresponds to the particle free
or bulk diffusivity in the absence of geometric restrictions and is
proportional to the temperature of the suspension fluid. We
numerically integrated Eq. (\ref{le}) by a Milstein algorithm
\cite{Milstein}. The stochastic averages reported in the forthcoming
sections were obtained as ensemble averages over 10$^6$ trajectories
with random initial conditions; transient effects were estimated and
subtracted. To check the reliability of our numerical simulations we solved the corresponding Fokker-Planck equation by means of a finite-element algorithm and obtained a good comparison with the Langevin method.

The driven particle current across the array strongly depends on the
force orientation. The problem cannot be reduced to a 1D problem,
unless ${\vec F}$ is oriented along a lattice axis, i.e.,
$\phi=\phi_{n,m}$ with $\phi_{n,m}=\arctan(nL_y/mL_{x})$ ($ n, m \in
\mathbb{Z}$ and $m \neq 0$). Due to its spatial symmetry, the array
can then be regarded as consisting of identical periodic channels
parallel to ${\vec F}$ and separated by reflecting walls. This allows to interpret the stationary transport properties of the array in terms of transport mechanisms through compartmentalized narrow (quasi 1D) channels. In Fig. \ref{F1} we illustrate the two limiting cases with $\phi=0$ and $\phi=\pi/4$ for a square lattice, $L_x=L_y$. For a rectangular lattice, i.e. $L_{x} \neq L_{y}$, and
$\phi=0$, the reduced channels are horizontal and have rectangular
compartments, $x^{_{\rm I}}_L\times y^{\rm I}_L$ with $x^{_{\rm I}}_L=L_{x}$ and $y^{_{\rm I}}_L=L_{y}$. For $\phi=\phi_{1,1}=\arctan{L_{y}/L_{x}}$, the reduced channels are diagonal and their compartments have width
$y^{_{\rm II}}_L=2L_{x}L_y/\sqrt{L_x^2 + L_y^2}$ and period $x^{_{\rm II}}_L=\sqrt{L_{x}^2 + L_{y}^2}$. For a square lattice, the latter channels coincide with the channel II of Fig. \ref{F1}(c), where $y_{L}^{_{\rm II}}=x_{L}^{_{\rm II}}= L_{x} \sqrt{2}$. Note that for symmetry reasons, the discussion of channels I and II can be further simplified by halving the channel width, see Figs. \ref{F1}(b) and (c).

Motivated by these observations we investigated Brownian transport in two categories of reduced periodic channels, conventionally directed along the $x$ axis (and parallel to $\vec F$). They correspond to cutting respectively the channel I of Fig. \ref{F1}(b) and the channel II of Fig. \ref{F1}(c) along the central dotted line and then varying the compartment length, $x_L$, at will (note that here and in what follows the superscripts are omitted). Channels I are thus characterized by a corrugated and a smooth wall, a geometry that lets straight particle trajectories through, no matter what the radius of the circular obstructions. Channels II have equally
corrugated walls, though shifted by half a period; for $r_0>y_L/4$
the driven particle can cross the channel only by circumventing the
obstructions on either walls. As a consequence, one expects distinct
transport properties for these two channel geometries
\cite{Bere1,Bere2,Lboro,PHfest,Borromeo}. Note that tuning the
compartment parameters corresponds to investigating rectangular
(channels I) and face-centered rectangular arrays in 2D (channels II) with ${\vec F}$ parallel to the principal axes.

Two transport quantifiers that best illustrate the different properties of channels I and II are the nonlinear mobility and the effective diffusion coefficient. We characterize the response of a Brownian particle dc-driven along the channel axis by computing its mobility $\mu$,
\begin{equation}
\mu(F) = \langle \dot x(F) \rangle /F,
\label{mobility}
\end{equation}
where $\langle \dot x(F) \rangle =\lim_{t\to\infty}[\langle x (t)
\rangle -x(0)]/t$, and its diffusivity $D$,
\begin{equation}
D(F)/D_0 = \lim_{t \to \infty}[\langle x^2(t)\rangle -
\langle x(t) \rangle^2]/2D_0t,
\label{diffusivity}
\end{equation}
both as functions of $F$ for different channel geometries. In the absence of external drives, Einstein's relation
\cite{Risken},
\begin{equation}
\mu_0 \equiv \mu(0) = D(0)/D_0,
\label{einstein}
\end{equation}
establishes the dependence of the transport quantifiers on the channel geometry and temperature under equilibrium conditions.

For a more compact presentation of our numerical data, we remark that Eq.~(\ref{le}) can be conveniently rewritten in terms of the rescaled units $t \to t D_0$ and $F \to F/D_0$. A straightforward dimensional argument shows that both the particle mobility, Eq. (\ref{mobility}), and its diffusivity in units of $D_0$, Eq. (\ref{diffusivity}), are functions of $F/D_0$, only, for any {\em given} channel geometry.

\begin{figure}[t]
\centering
\includegraphics[width=0.42\textwidth]{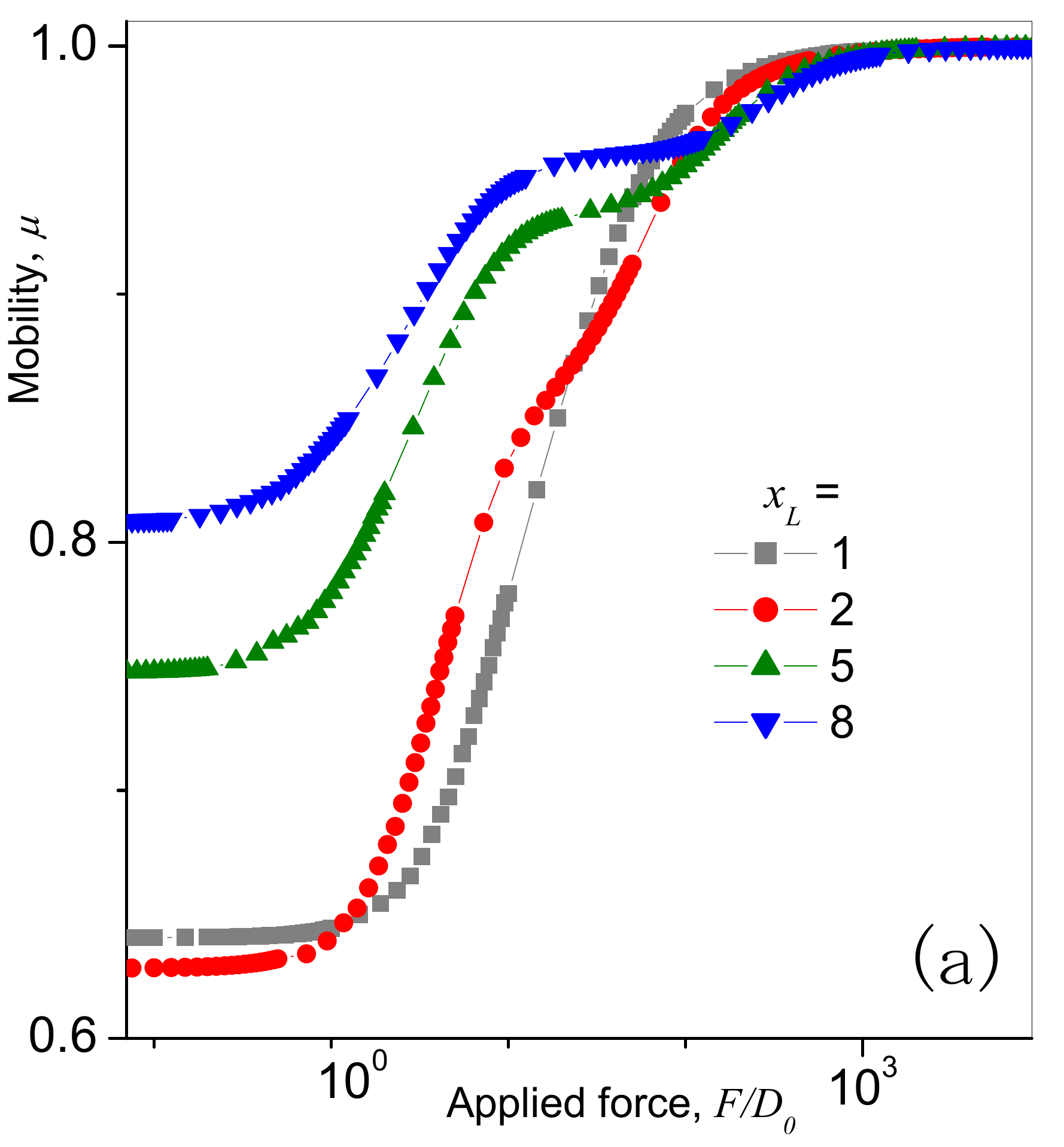}
\includegraphics[width=0.41\textwidth]{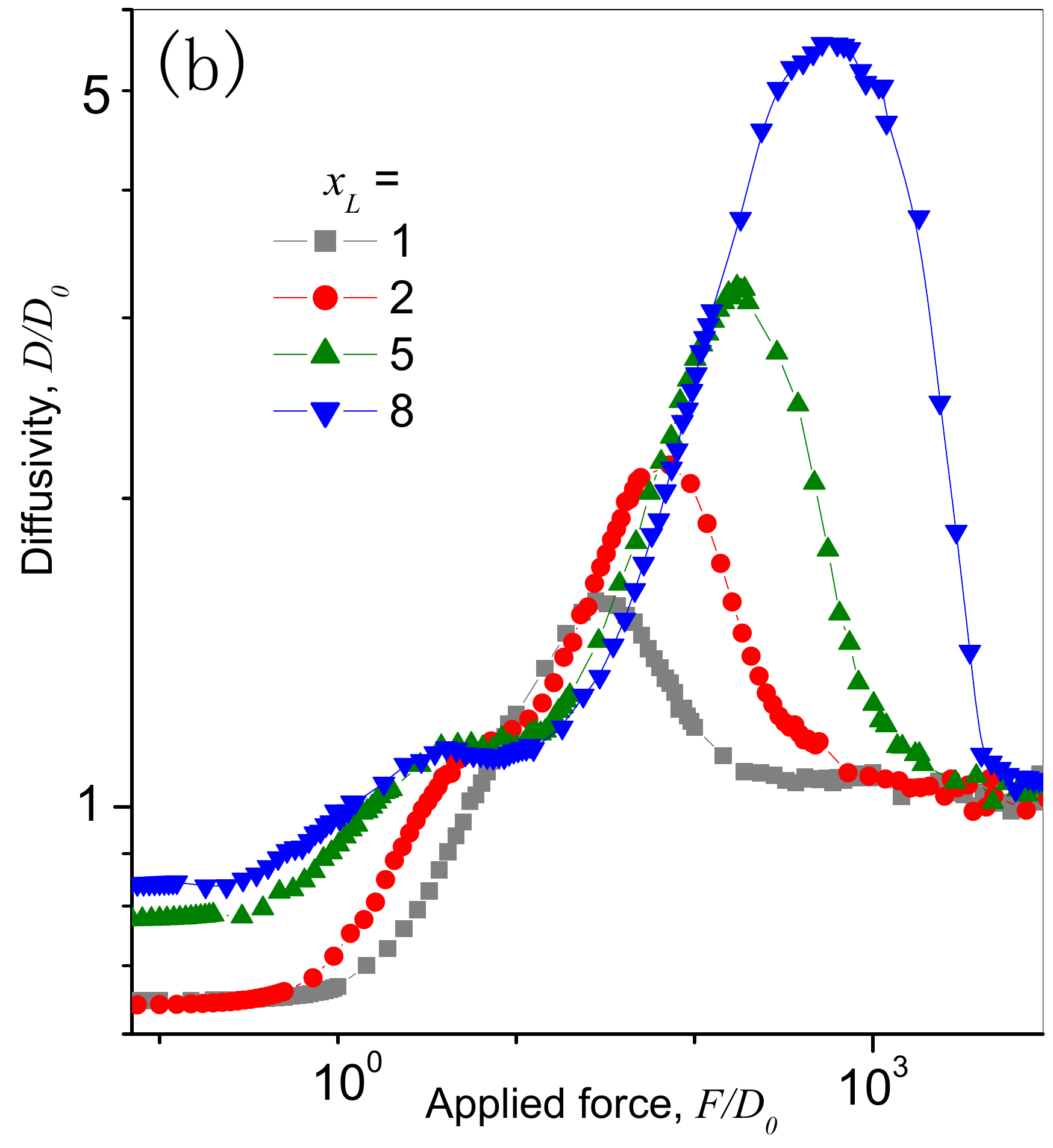}
\caption{(Color online) Driven transport in a channel I: particle mobility $\mu$ (a) and scaled diffusivity $D/D_0$ (b) vs. the scaled force $F/D_0$ for different $x_L$. Other simulation parameters are: $y_L=1$, $D_0=1$, and $r_0=0.4$.} \label{F3}
\end{figure}

\section{Channel transport}
\label{channels}

Contrary to smoothly corrugated channels, also called entropic
channels \cite{chemphyschem}, introduced first in Ref. \cite{Zwanzig}
and further investigated in
Refs. \cite{Reguera:2001,Kalinay,Laachi,Reguera:2006,Burada}, strongly
compartmentalized channels with narrow sharp bottlenecks, or pores,
cannot be analyzed in terms of an effective 1D kinetic process
directed along their axis \cite{Lboro}. Accordingly, driven transport
in such strongly constrained geometries exhibits distinct features,
which cannot be reduced to known properties of Brownian motion in 1D
periodic systems \cite{chemphyschem,Risken}.

We outline here the main differences between transport in entropic and
sharply compartmentalized channels. In entropic channels $\mu(F)$
{\it increases} from a relatively small value for $F=0$, $\mu_0$ of
Eq. (\ref{einstein}), up to the free-particle limit, $\mu_\infty=1$,
for $F\to \infty$ \cite{Reguera:2006,Burada}. On the contrary, in coaxial
compartmentalized channels $\mu(F)$ {\it decreases} monotonically with
increasing $F$ towards a geometry-dependent asymptotic value,
$\mu_\infty$, equal to the ratio of the pore to the channel
cross-section \cite{Lboro}. Such an asymptote vanishes for eccentric
compartmentalized channels with off-axis, non-overlapping pores
\cite{Borromeo}.

Significant differences have also been reported for the
diffusivity. For entropic channels with smooth pores, the function
$D(F)$ approaches the free-diffusion limit for $F\to \infty$,
$D(\infty)=D_0$, after going through an excess diffusion peak centered
at an intermediate (temperature dependent \cite{Reguera:2006,Burada}) value of the drive. Such a peak signals the depinning of the particle from the
entropic barrier array \cite{Costantini}. In coaxial compartmentalized channels, instead, $D(F)$ exhibits a distinct quadratic dependence on
$F$ \cite{Bere2,PHfest}, reminiscent of Taylor's diffusion in
hydrodynamics. This observation suggests that the particle never frees itself from the geometric constriction of the compartment pores, no matter how strong $F$. The quadratic divergence $D(F)$ can be cut off for suitably large $F$ by alternately shifting the pores off-axis (eccentric channels \cite{Borromeo}).

Driven transport in channels I and II of Fig. \ref{F1} is characterized by the superposition of, or sometimes the competition between, the properties observed for the two opposite compartmentalization geometries considered in the earlier literature. We remind here that, due to the mirror symmetry of the obstacle geometries of Figs. \ref{F1}(b) and (c), our analysis refers  to the irreducible channels I and II obtained by cutting the channels sketched there along their axis (dotted line) and then varying their spatial periodicity. Therefore, the irreducible channels discussed below have effective width $y_L/2$ and transverse bottleneck width $\Delta=y_L/2-r_0$.

\begin{figure}[t]
\centering
\includegraphics[width=0.44\textwidth]{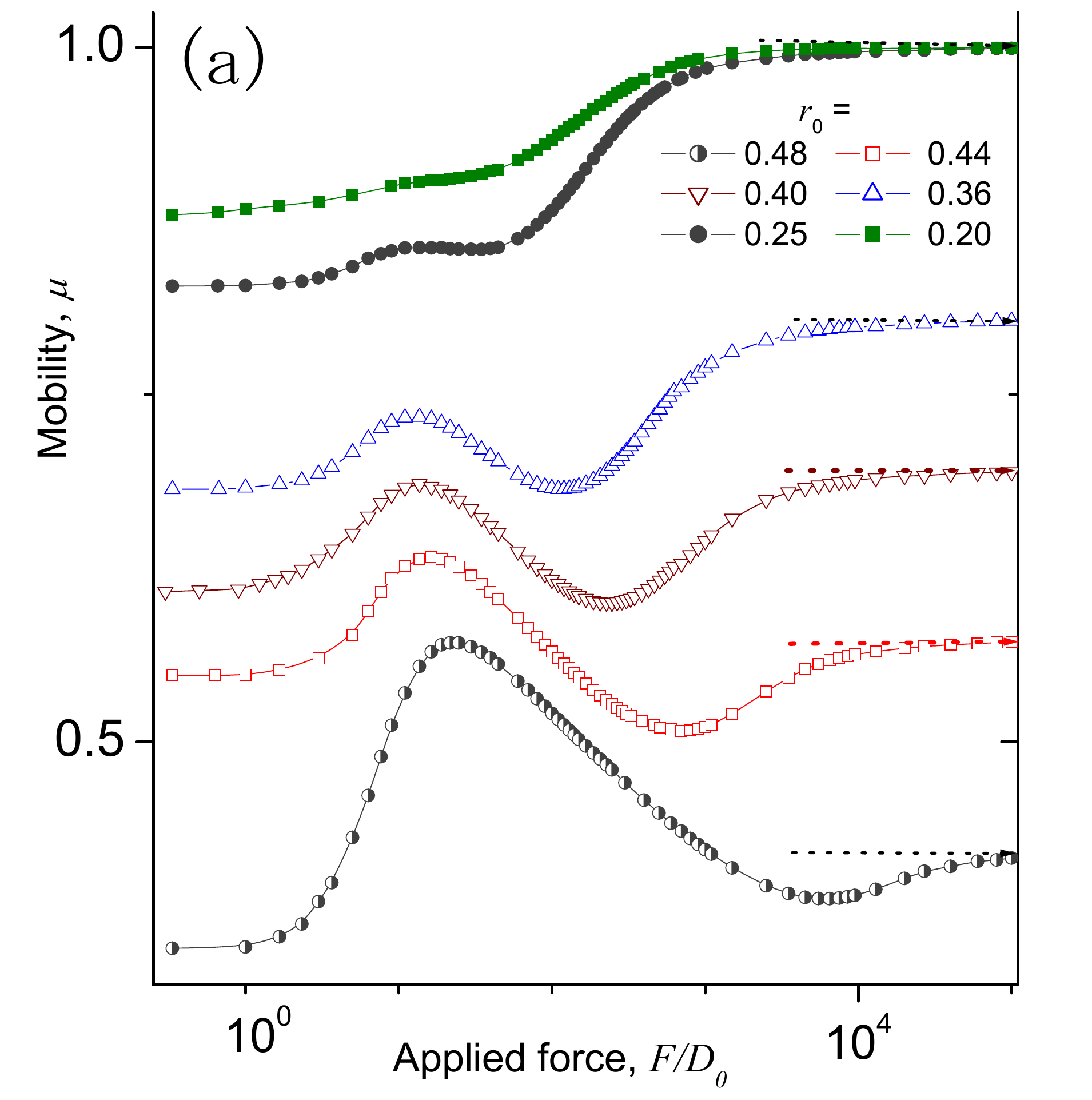}
\includegraphics[width=0.42\textwidth]{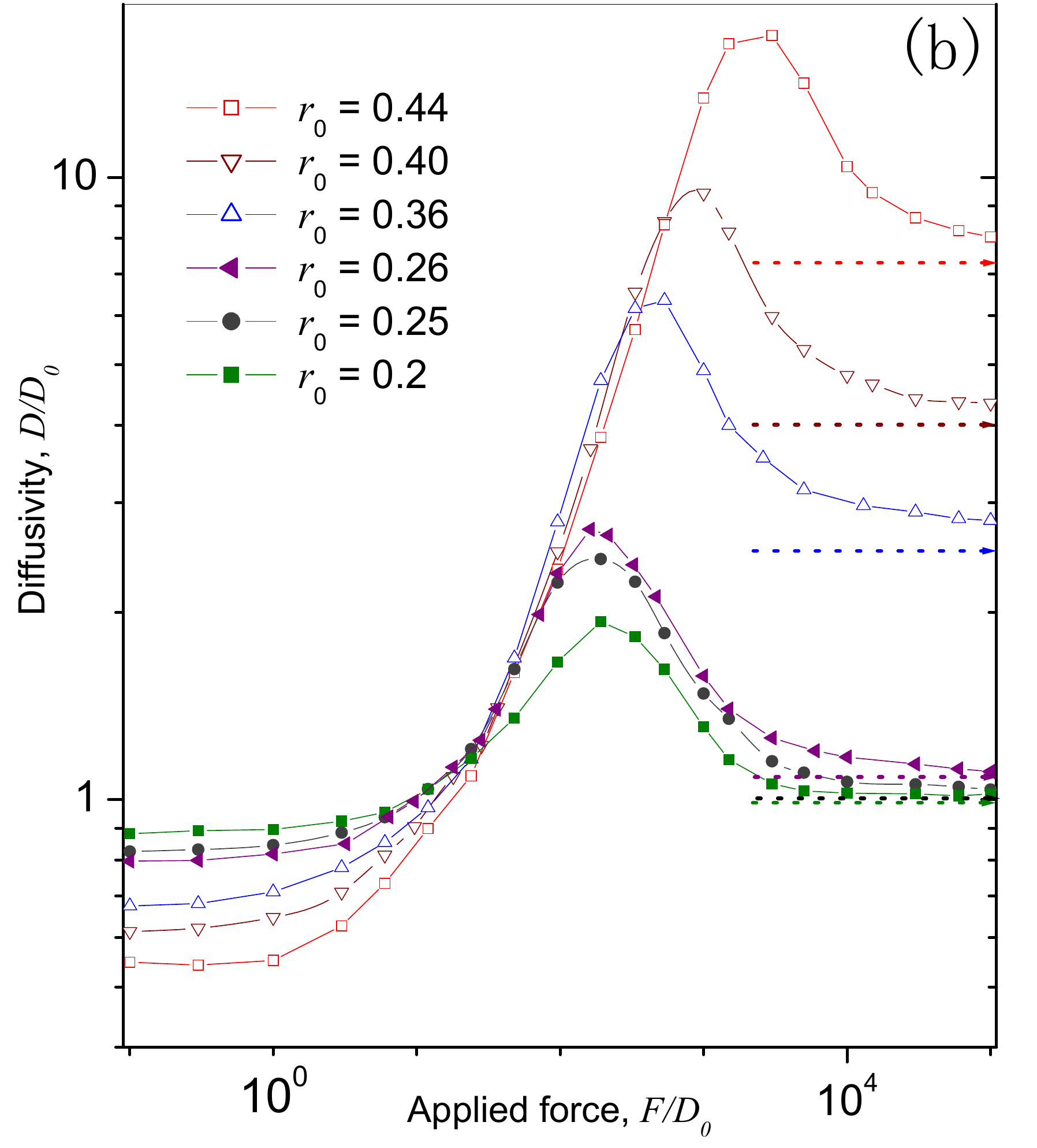}
\caption{(Color online) Driven transport in a channel II with $x_L=2$, $y_L=1$, $D_0=0.03$, and different $r_0$: particle mobility $\mu$ (a) and scaled diffusivity $D/D_0$ (b) vs. the scaled force $F/D_0$. Dotted lines represent $\mu_\infty$, Eq. (\ref{muOO}), in (a) and $D_{\infty}$, Eq. (\ref{DOO}), in (b).} \label{F4}
\end{figure}

\subsection{Channels I}
\label{chI}
Channels I corresponding to square arrays, $x_L=y_L$, exhibit transport properties apparently not much different from the entropic
channels. In particular, the mobility is an increasing function of
$F$; the $\mu(F)$ curves jump from $\mu_0$ to $\mu_\infty=1$ sharper
and sharper as the radius $r_{0}$ of the obstacles is increased [Fig. \ref{F2}(a)]. The diffusivity curves develop an excess diffusion peak in correspondence with the mobility surge; the height of these peak weakly depend on $r_{0}$.

New features start to emerge when we consider channels with
rectangular compartments, i.e. $x_{L}>y_{L}$. These become apparent in Fig. \ref{F3}, where we display data for simulated transport in
channels I for different period lengths $x_L$. The occurrence of
two distinct diffusion mechanisms is proved by the two-step
increase of the mobility and the corresponding double diffusivity
peaks. Such peaks merge as $x_L$ tends to $y_L$.

\subsection{Channels II}
\label{chII}
In Fig. \ref{F4} we plot the curves $\mu(F)$ and $D(F)$ for
channels II with comparable compartment dimensions, $x_L$ and $y_L$, and different obstacle radius, $r_0$. The mobility function, $\mu(F)$, develops a non-monotonic dependence on $F$, which
becomes prominent for narrow bottleneck widths, $\Delta=y_{L}/2-r_{0}$, but vanishes altogether for $\Delta\geq y_L/4$ [Fig. \ref{F4}(a)]. Negative differential mobility is a peculiar feature of all sharp compartmentalized channels \cite{Lboro} and eccentric channels, in particular \cite{Borromeo}. Indeed, each compartment of the reduced channels II of Fig. \ref{F2}(c) has two off-axis bottlenecks, located alternately against its top and  bottom wall. The subsequent mobility surge toward a horizontal asymptote with $\mu_\infty < 1$
indicates that on increasing $F$ the particle grows more sensitive to the pinning action of the obstacles, which it overcomes initially by
mere Brownian diffusion and eventually with the assistance of the
drive itself.

Correspondingly, such change in the depinning mechanism is signaled by a conspicuous diffusion peak [Fig. \ref{F4}(b)]. In comparison to Fig. \ref{F2}(b), here the $D(F)$ peaks are more than one order of magnitude larger than $D_0$ and their position strongly depends on
$\Delta$ (or $r_0$). Most remarkably, the asymptote $D(\infty)$ is
systematically larger than the free diffusivity, $D_0$, for $\Delta
\leq y_L/4$, like for eccentric compartmentalized channels
\cite{Borromeo}. Note also that the slowly increasing branches of the curves plotted in Fig. \ref{F4}(b) tend to overlap and grow slower
than for coaxial compartmentalized channels, where $D(F)$ diverges like $F^2$ \cite{Bere2,Borromeo}.

\begin{figure}[t]
\centering
\includegraphics[width=0.4\textwidth]{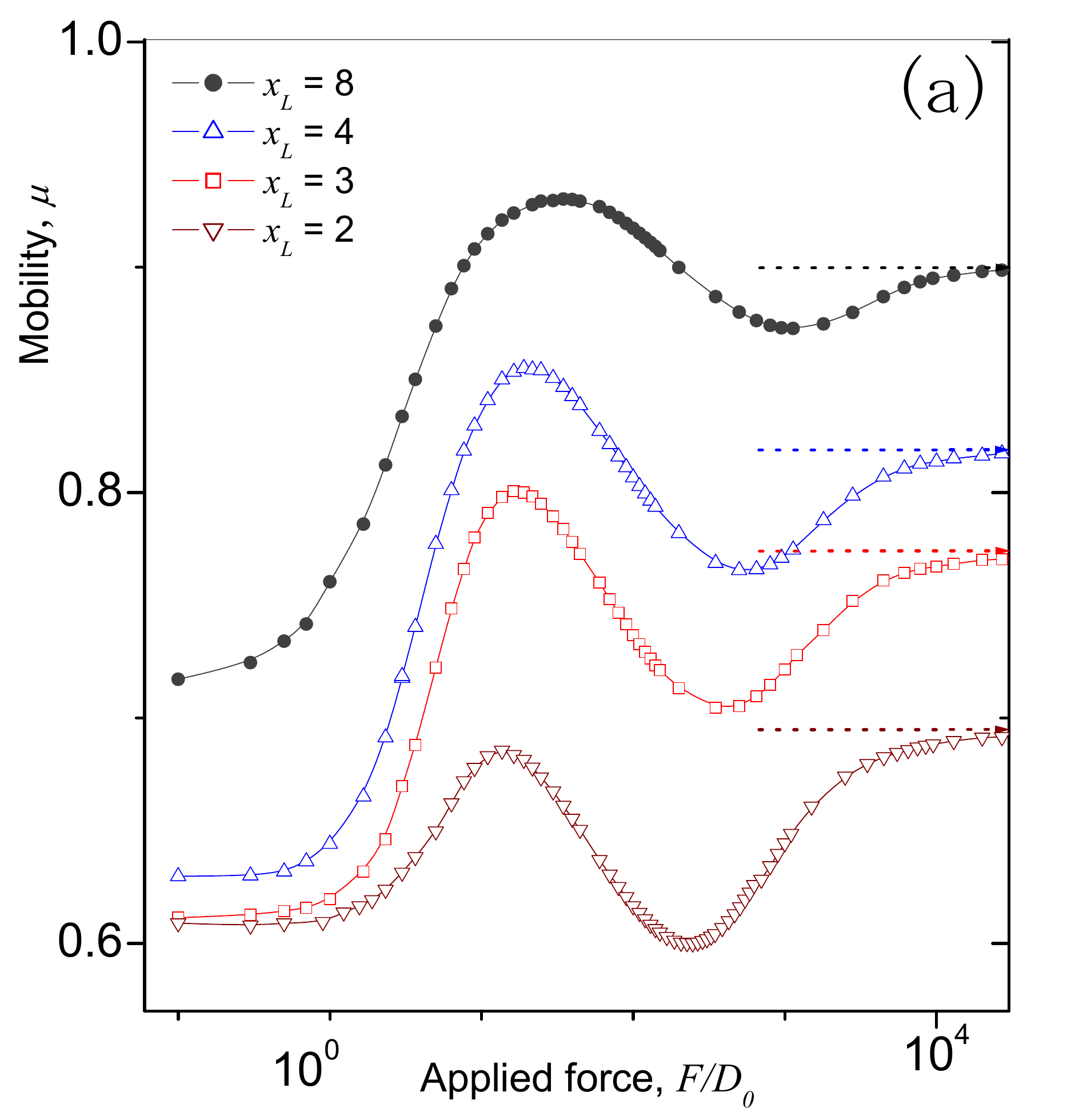}
\includegraphics[width=0.4\textwidth]{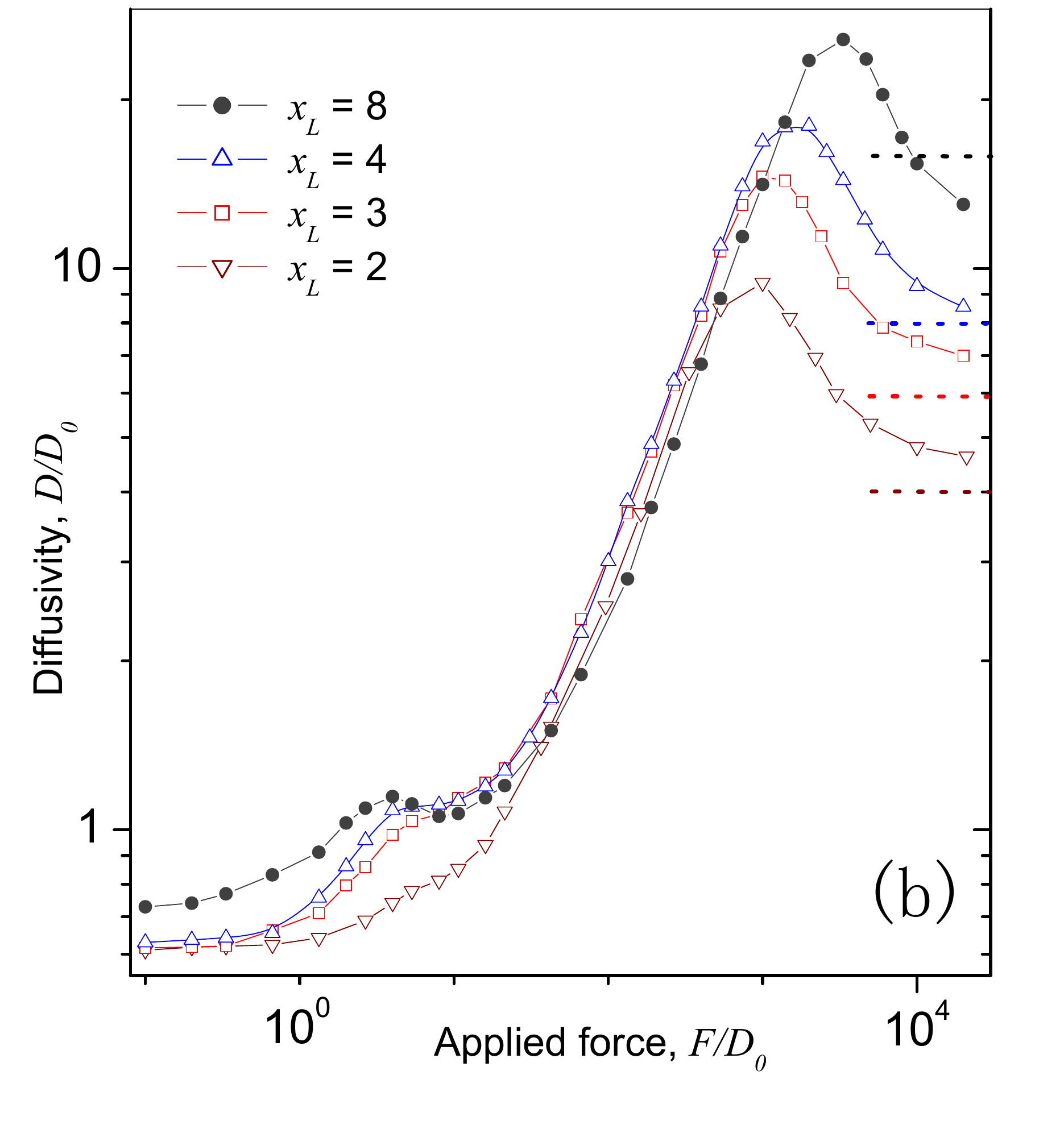}
\caption{(Color online) Driven transport in a channel II with $y_L=1$, $D_0=0.03$, $r_0=0.4$ and different period $x_L$: particle mobility $\mu$ (a) and scaled diffusivity $D/D_0$ (b) vs. the scaled force $F/D_0$. Dotted lines represent $\mu_\infty$, Eq. (\ref{muOO}), in (a) and $D_{\infty}$, Eq. (\ref{DOO}), in (b).} \label{F5}
\end{figure}

Like for channels I, two different transport mechanisms
became apparent as we space the circular obstacles out along the
channel axis, i.e., for $x_L >> y_L$. The diffusivity curve $D(F)$ develops a secondary peak
at lower $F$, which grows prominent on further increasing $x_L$
[Fig. \ref{F5}(b)]. At variance with the high-$F$ peak, the position
of the emerging peak is rather insensitive to $x_L$. Correspondingly, the mobility curves shift toward higher values, their most apparent
feature remaining its non-monotonic $F$ dependence. By closer
inspection, for large $x_L$, in correspondence with the secondary
$D(F)$ peak one notices the appearance of a shoulder, or step, on the low-$F$ raising branch of $\mu(F)$.

\section{Depinning mechanisms}
\label{theory}

\begin{figure}[t]
\centering
\includegraphics[width=0.45\textwidth]{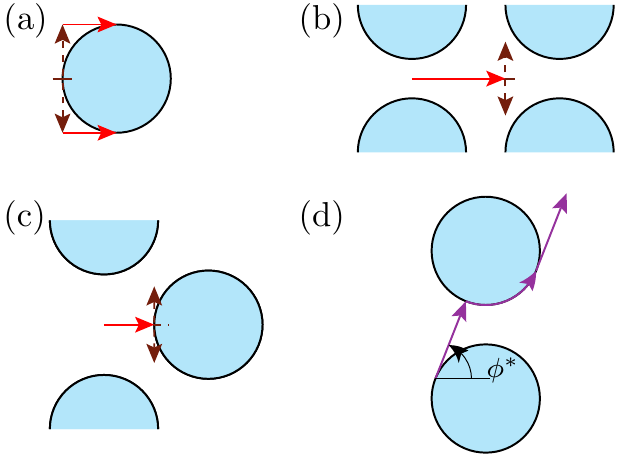}
\caption{(Color online) Depinning mechanisms. (a) The Brownian particle circumvents a single obstacle by diffusing in the transverse direction (vertical arrows) and then drifting along the axis driven by $F$ (horizontal arrow). (b),(c) Trajectory focusing along the channel lane results from the competing action of longitudinal drift (horizontal arrows) and transverse diffusion (vertical arrows) across the array. The transverse diffusion length is of the order of half the channel width in (b) and half the bottleneck width in (c). (d) Example of minimum longitudinal mobility orientation in a rectangular 2D array, see Sec. \ref{angular}.} \label{F6}
\end{figure}

The most remarkable property of transport in channels I and II emerges
for rectangular compartments with $x_L >y_L$. As anticipated in
Secs. \ref{chI} and \ref{chII}, the two-peaked structure of $D(F)$
results from the competition between the two depinning mechanisms
illustrated in Fig. \ref{F6}. The Brownian particle overcomes the
obstacle thanks to the combined action of the drive $F$ and the noise
$\xi(t)$. The drive, in particular, on one side pushes the particle
against the obstacles, thus enhancing the low-$F$ pinning by the array
(raising $D(F)$ branch), on the other side guides the particle around
the obstacles, so as to collimate its trajectories through a coaxial
[channels I, $D(\infty)=D_0$] or eccentric sequences of bottlenecks
[channels II, $D(\infty)>D_0$].

For relatively low drives, the particle circumvents a single circular obstacle, irrespective of the array geometry and the drive orientation, by {\em diffusing} a transverse distance $r_0$ over the time it takes to {\em drift} the same distance $r_0$ [Fig. \ref{F6}(a)]. On equating the transverse diffusing time, $\tau_{\perp}=r_0^2/2D_0$, and the longitudinal drift time, $\tau_{\parallel}=r_0/F$, we locate the low-$F$ depinning threshold at
\begin{equation}\label{F_1}
F^{(1)}/{D_0}=2/r_0.
\end{equation}
We recall that excess diffusion peaks are characteristic signatures of depinning thresholds \cite{BM1,Costantini}. We stress that this estimate for $F^{(1)}$ is expected to apply to both types of channels, I and II, and to any geometry and size of their compartments (i.e., to any 2D array). As shown below, in the case of square compartments, Figs. \ref{F2}(b) and \ref{F4}(b), low- and high-$F$ peaks cannot be separated. The estimate of Eq. (\ref{F_1}) closely locates the low-$F$ peaks of all $D(F)$ curves reported in Figs. \ref{F3}(b) and \ref{F5}(b). Note that in the limit of narrow bottlenecks, $\Delta \to 0$, $F^{(1)}/{D_0}$ approaches $4/y_L$, irrespective of $x_L$ [Fig. \ref{F5}(b)].

For large drives, channels I and II behave differently. The
bottlenecks in channels I are coaxial [Fig. \ref{F6}(b)]. This means
that the driven Brownian particle gets trapped within a couple of
obstacles if, during the {\em drift} time it takes to cover the
distance separating the center of a bottleneck from the nearest pair
of obstacles, $x_L-r_0$, it {\em diffuses} a transverse distance of
the order of half the width of the (reduced) channel, $y_L/4$. Equating the
corresponding times, $\tau_{\perp}=y_L^2/32D_0$ and
$\tau_{\parallel}=(x_L-r_0)/F$, we obtain an estimate for the position of the high-$F$ diffusivity peaks in channels I, namely,
\begin{equation}\label{F_2I}
F^{(2,\rm I)}/{D_0}=32(x_L-r_0)/y_{L}^2\, .
\end{equation}

In channels II with $\Delta < y_L/4$, i.e. $r_{0}>y_{L}/4$, the focusing action of the drive forces the particle through a meandering path. This action becomes effective when the particle {\em drifts} the distance separating a bottleneck from the opposing obstacle [Fig. \ref{F6}(c)], $x_L/2-r_0$, while {\em diffusing} across the lane passing through the bottleneck of a (reduced) channel, i.e., a transverse distance of the order of half its width, $\Delta/2$. The corresponding depinning threshold is thus estimated to be
\begin{equation}\label{F_2II}
F^{(2,\rm II)}/{D_0}=4(x_L-2r_0)/\Delta^2.
\end{equation}

Contrary to $F^{(1)}$, the thresholds $F^{(2,\rm I)}$ and $F^{(2,\rm II)}$
depend on the channel geometry, being different for channels I and II, and linearly shift to higher values with increasing the length of the channel compartments, $x_L$, or shrinking the obstacle size, $r_0$.  Despite the rough estimate of the characteristic transverse
diffusing length, the thresholds of Eqs. (\ref{F_2I}) and (\ref{F_2II}) locate quite closely the large drive diffusion peaks in panels (b) of Figs. \ref{F2}-\ref{F5}. For square channel compartments, $x_L=y_L$, with narrow bottlenecks, $r_0\simeq y_L/2$, the two depinning mechanisms are not clearly distinguishable, being the size of the obstacle and the array lattice constant of the same order. Correspondingly, the diffusivity curves exhibit only one peak.

The limiting values of $\mu(F)$ and $D(F)$ for $F=0$ and $F\to \infty$ are also of some interest. For $F=0$, mobility and diffusivity are related through Einstein's identity, Eq. \ref{einstein}. Analytical expression for $\mu_0$ in restricted geometries can be obtained only under special conditions \cite{Bere1,PHfest}. For instance, in channels I and II corresponding to {\em square} arrays with narrow bottlenecks, $\Delta \to 0$, $\mu_0$ can be expressed in terms of the mean first exit time for the Brownian particle to escape the interstitial region delimited by four nearest neighboring obstacles [Fig. \ref{F2}(b), inset]. Our numerical simulations show that, as expected, the mean exit time through all four openings, $\tau_4$, is half the mean exit time through any pair of opposite openings, $\tau_2$, with both $\tau_2$ and $\tau_4$ decaying inversely proportional to the pore width, i.e., like $1/\sqrt{\Delta}$. These results are well fitted by the law,
\begin{equation}\label{tau2}
\tau_2=2\tau_4\simeq\left ( 1-\frac{\pi}{4}\right )\frac{x_L^2 \pi}{8D_0\sqrt{\Delta/x_L}},
\end{equation}
which is consistent with a recent analytical prediction [see Eq. (13) of Ref. \cite{Schuss}]. Moreover, the mean sojourn time in a square
channel I compartment is $2\tau_2$ and, accordingly,
$\mu_0=x_L^2/4D_0\tau_2$, also in close agreement with the numerical
data for small $\Delta$ plotted in Fig. \ref{F2}. This argument can be easily generalized to rectangular arrays, as long as the distance between neighboring obstacles remains sufficiently small compared to the lattice constants.

The asymptotes $\mu(\infty)$ and $D(\infty)$ in channels II with $\Delta <y_L/4$ also deserve attention. In the limit $F\to \infty$, the pore eccentricity, which is responsible for the {\em negative differential mobility} of channels II, is eventually superseded by the funneling effect due to the rounded shape of the obstacles. In such limit, the particle trajectories consist of straight parallel segments, which a free particle traverses with speed $F$, and arcs of the obstacle boundaries, as shown in Fig. \ref{F1}(c), where the driven particle slows down. On averaging out noise fluctuations, the total time taken by the particle to cross one channel compartment, $\tau_c$, can be derived analytically, so that, $\mu_\infty=x_L/\tau_cF$, that is
\begin{equation}\label{muOO}
\mu_\infty=\left [1-\frac{y_L}{x_L}\sqrt{2\epsilon-1}+\frac{r_0}{x_L}\ln \frac{\epsilon+\sqrt{2\epsilon-1}}{\epsilon-\sqrt{2\epsilon-1}}\right ]^{-1},
\end{equation}
with $\epsilon=2r_0/y_L$ and $1/2 < \epsilon < 1$. The predicted values for $\mu_\infty$ are indicated by dotted lines in Figs. \ref{F4}(a) and \ref{F5}(a). Note that in channels II for $\Delta \geq y_L/4$ and in all channels I the traversal time is $\tau_c=x_L/F$ and, therefore, $\mu_\infty=1$.

A full analytic calculation of $D(\infty)$ proved to be a cumbersome task. As in channels II each bottleneck is faced by a blocking obstacle, one would expect that for $\Delta <r_0$ the diffusivity $D(\infty)$ scales like $r_0/\Delta$, namely the ratio of the obstacle diameter to the vertical distance between two neighboring obstacles. Furthermore, on regarding the channel as a 1D discretized structure with spatial constant $x_L$, one can formally express $D(\infty)$ as $x_L^2/2{\bar \tau}$, where $\bar \tau$ is a certain diffusion time constant, which we know to be proportional to the compartment volume \cite{Schuss}. Hence, even without calculating $\bar \tau$, one concludes that $D(\infty)$ is proportional to $x_L$. Therefore, we tried to reproduce our numerical data for $D(\infty)$ by means of the heuristic law
\begin{equation}\label{DOO}
\frac{D(\infty)}{D_0}=\frac{x_L}{2y_L}~\frac{r_0}{\Delta},
\end{equation}
which turns out to work well both in Fig. \ref{F4}(b) and in Fig. \ref{F5}(b). Note that, in the case of larger $x_L$, Fig. \ref{F5}(b), where Eq. (\ref{DOO}) seems to work not as well, a conclusive estimate of the asymptote $D(\infty)/D_0$ would require simulating even higher $F/D_0$ values, which in practice becomes very difficult to implement numerically because this regime would require exceedingly small time steps.

\begin{figure}[t]
\centering
\includegraphics[width=0.35\textwidth]{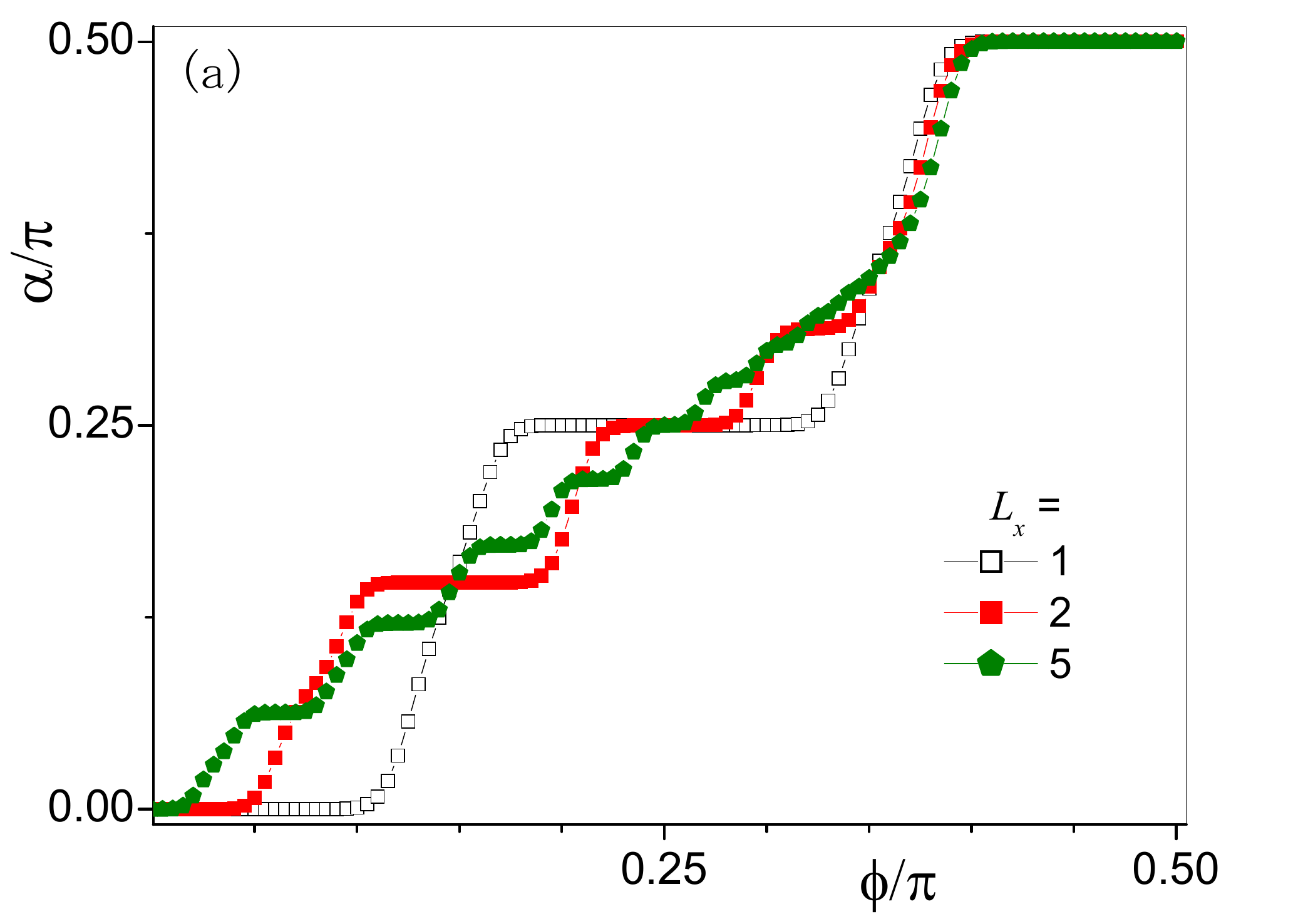}
\includegraphics[width=0.36\textwidth]{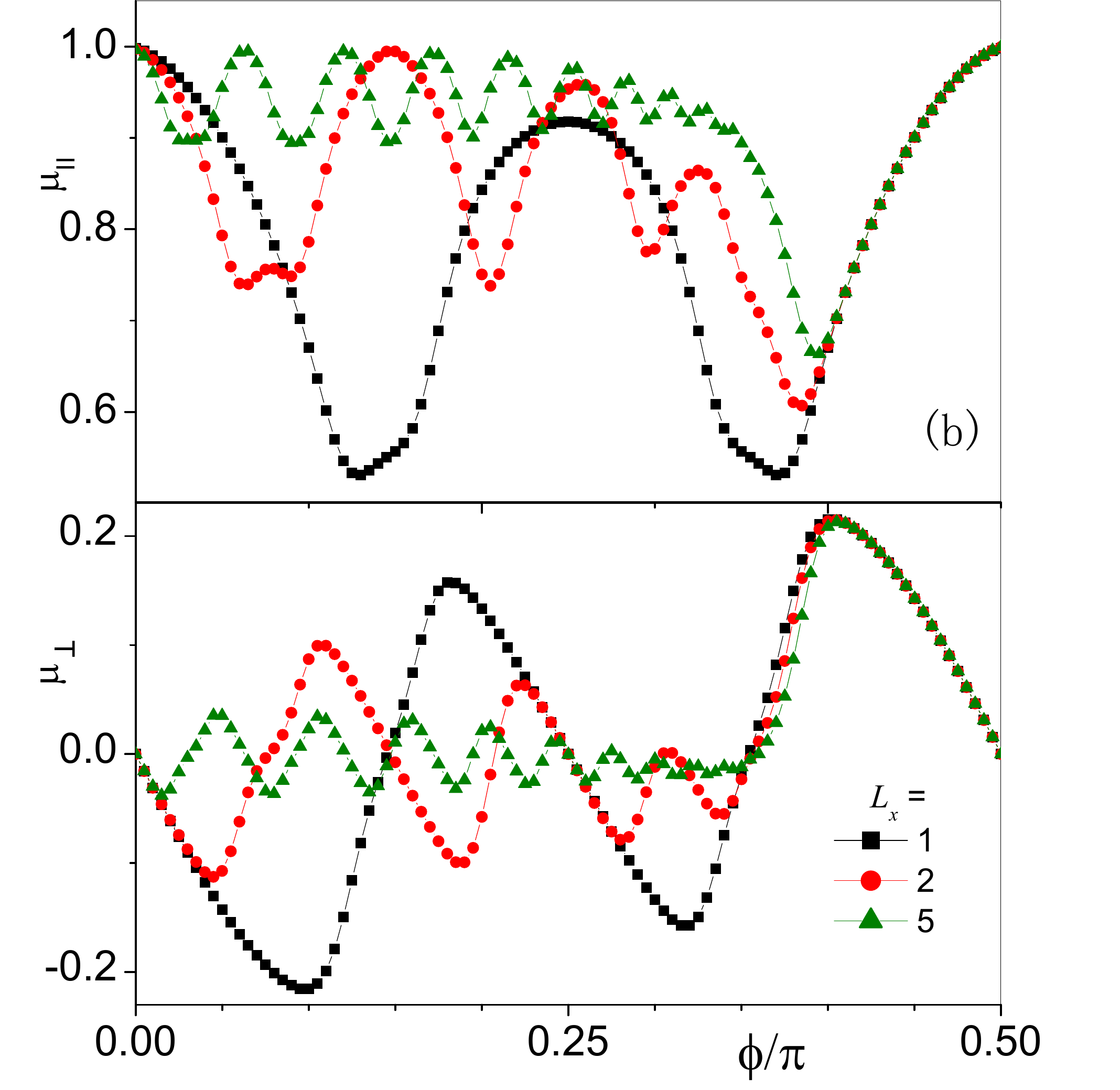}
\includegraphics[width=0.36\textwidth]{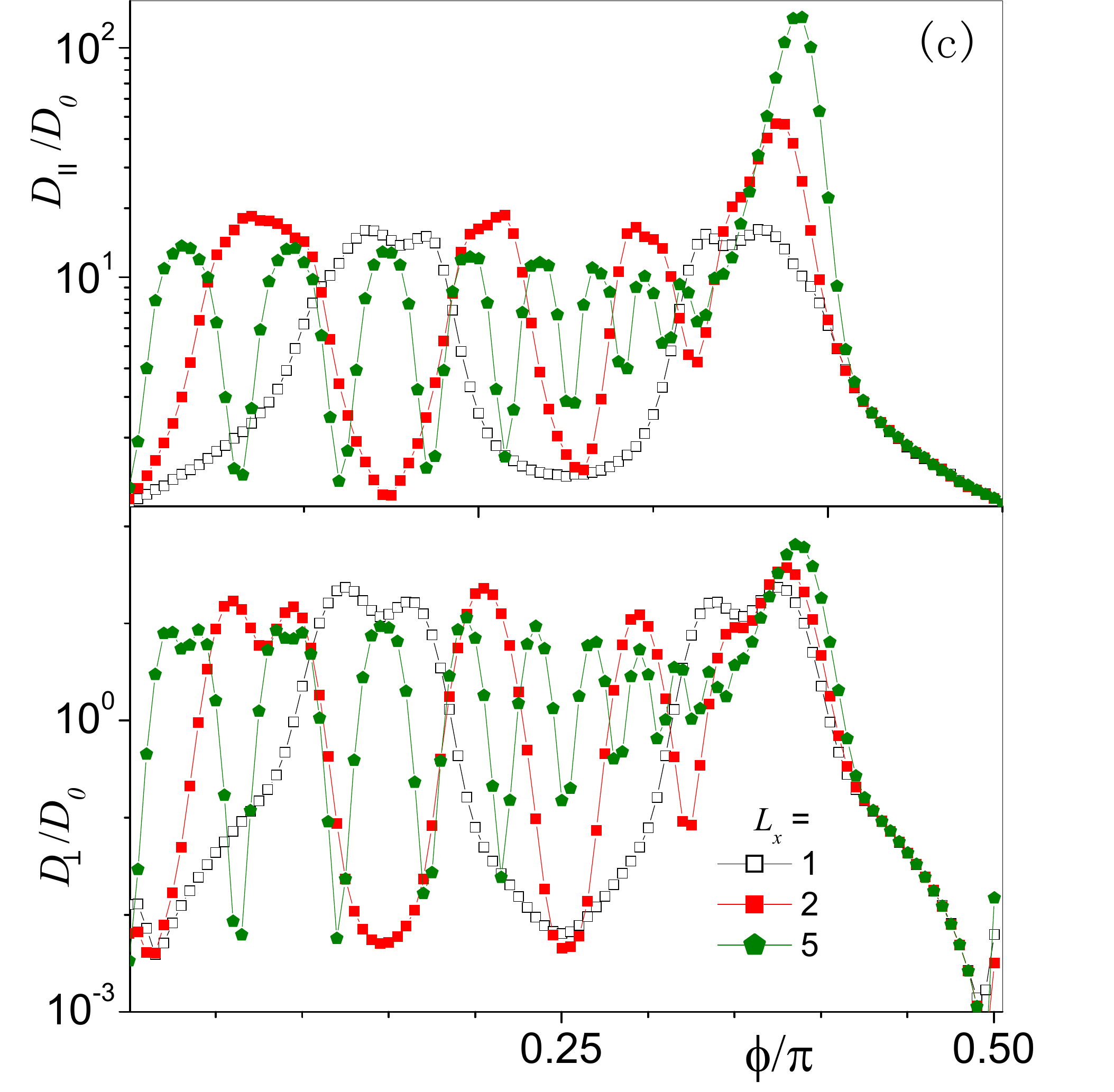}
\caption{(Color online) Angular effects. (a) Migration angle; (b) longitudinal, $\mu_{\perp}$, and lateral mobility, $\mu_{\parallel}$; and (c) scaled longitudinal, $D_{\perp}/D_0$, and lateral diffusivity, $D_{\parallel}/D_0$, of a Brownian particle driven at an angle $\phi$ in rectangular arrays with different $L_x$ (see legends). All quantities are plotted vs. $\phi$ for $L_y=1$, $r_0=0.4$, and $F/D_0=10^3$. In these simulations we set $D_0=1$.} \label{F7}
\end{figure}

\section{Parallel and lateral transport}\label{angular}

We address now the practical situation
\cite{Brenner,Tierno,MacDonald,Huang,Nori,Riken,Moshchalkov,Drazer1,Drazer2,Drazer3,Souto}
when the orientation of the drive can be varied at will with respect to the array principal axes. The angle $\phi$ in Fig. \ref{F1}(a) can thus be rotated between $0$ and $\pi/2$. For a square lattice, $L_x=L_y$, this corresponds to continuously switch from a channel I ($\phi=0$) to a channel II ($\phi=\pi/4$) and then back again to the initial channel I geometry ($\phi=\pi/2$). For a rectangular 2D lattice, $L_x\neq L_y$, the two channels I have different width, that is, $L_y$ for $\phi=0$ and $L_x$ for $\phi=\pi/2$, while no channel
II is recovered. Indeed, as ${\vec F}$ is directed along a diagonal of the rectangular lattice cell, the corresponding reduced 1D channel looks similar to channel II in Fig. \ref{F1}(c), but for a relative shift of the lower and upper obstacle rows. The transport properties of such diagonal channels are qualitatively the same as for channels II of Sec. \ref{chII}.

In Fig. \ref{F7} we display some significant simulation data for driven Brownian transport in square and rectangular arrays. To this purpose we generalized the definitions of Eqs. (\ref{mobility}) and (\ref{diffusivity}) to extract the mobility and the diffusivity in the direction parallel and orthogonal to ${\vec F}$. The corresponding four transport quantifiers are denoted by $\mu_{\parallel}$, $D_{\parallel}$ for the longitudinal transport and by $\mu_{\perp}$, $D_{\perp}$ for the lateral or transverse transport.

The salient properties of directed transport in 2D arrays are easily reconciled to the geometric properties of the particular system at hand. With focus on the $\phi$ dependence of the transport quantifiers, one notices that:

{\it(i)} For square arrays and vanishingly small drives, $D_{\perp}(0)=D_{\parallel}(0)$ and  $\mu_{\perp}(0)=\mu_{\parallel}(0)$, irrespective of $\phi$. This is consistent with the $D_4$ symmetry of the array lattice and with the fact that the mobility and diffusivity components are defined in the long-time limit. Applying a finite drive to a square array breaks the $D_4$ symmetry and leads to weaker angular symmetry relations, $D_{\parallel,\perp}(\psi)=D_{\parallel,\perp}(-\psi)$, $\mu_{\parallel}(\psi)=\mu_{\parallel}(-\psi),$ and $\mu_{\perp}(\psi)=-\mu_{\perp}(-\psi),$ where $\psi=\phi-\pi/4$. Note that the lateral mobility is an odd function of $\psi$, all other transport quantifiers being even functions.\\
{\it(ii)} Driven longitudinal transport in a {\it square} array is the least efficient for two special orientations, $\pm\psi^*$, or $\phi^*$ and $\pi/2-\phi^*$, as signaled by the occurrence of two symmetric dips in $\mu_{\parallel}$ and peaks in $D_{\parallel}$ and $D_{\perp}$ [Figs. \ref{F7}(b) and (c)]. The lateral mobility, $\mu_{\perp}$, vanishes for $\phi=0, \pi/2$ (channel I geometry) and in correspondence with the diffusion peaks, $\psi=\pm \psi^*$. The interpretation of such angular dependence follows from the argument leading to Eq. (\ref{muOO}). A noiseless driven trajectory consists of straight segments parallel to ${\vec F}$ and circular arcs running around the obstacles.  Each straight segment is tangent to the circumvented obstacle and impinges on the next obstacle that stands on the particle's way. The blocking action of this second obstacle is the largest, i.e., $\mu_{\parallel}$ has a minimum, when the incident segment runs along one of its diameters [Fig. \ref{F6}(d)]. Under these circumstances, the trajectory can run around the blocking obstacle on either side with equal probability, i.e., $\mu_{\perp}=0$, and the dispersion of the transport current in both directions, i.e. $D_{\parallel}$ and $D_{\perp}$, is maximum. For narrow bottlenecks a simple calculation yields $\phi^*=\arccos \sqrt{1-(r_0/L_y)^2}$, in agreement with the plots of Fig. \ref{F7}.\\
{\it(iii)} The lateral diffusivity in driven arrays is suppressed for increasing $F$ (not shown). This effect signals that the particle is effectively channeled by oriented rows of bottlenecks [possibly at an angle with ${\vec F}$, see Fig. \ref{F7}(a)]. Such mechanism sets on when the driven particle crosses an array unit cell in a time much shorter than the time it takes to diffuse across a single bottleneck. 
The effective migration angle \cite{Drazer3}, $\alpha$, of a particle driven across a 2D array does not necessarily coincide with the drive orientation. In fact, $\alpha=\phi+\arctan(\mu_{\perp}/\mu_{\parallel})$. This means that $\alpha$ and $\phi$ coincide only for $\mu_{\perp}=0$, that is for $\psi=\pm \psi^*$ and $\pm \pi/4$. For a square lattice the angular dependence of $\alpha(\phi)$ is plotted in Fig. \ref{F7}(a). Due to the angular symmetry of $\mu_{\perp}$ and $\mu_{\parallel}$, it follows that $\alpha(\psi)+\alpha(-\psi)=\pi/2$. The step-like structure of $\alpha(\phi)$ is controlled by the commensuration of the lattice constants, as first reported in Ref. \cite{Drazer3}.
\\
{\it(iv)} For rectangular arrays, the angular symmetries of the transport quantifiers in (i)-(iii) are broken. The nodes of the curves $\mu_{\perp}(\phi)$ plotted in Fig. \ref{F7}(b) increase in number with the ratio $L_x/L_y$. Indeed, reflecting the lower rotational lattice symmetry, there exists more distinct critical angles, $\phi^*$, that satisfy the maximal blocking condition of item (ii). However, as $L_x$ grows much longer than $L_y$, the lateral mobility for $\phi<\pi/4$ gets suppressed, because channeling in the horizontal direction becomes more effective than in the vertical direction. Correspondingly, the migration angle $\alpha(\phi)$ develops more steps. For commensurate lattice constants, i.e., $L_x/L_y=n$ with $n$ an integer, we counted exactly $n$ steps of the $\alpha(\phi)$ curves within the interval $0<\phi\leq \pi/4$, each corresponding to a node of $\mu_{\perp}(\phi)$. For $\pi/4 < \phi\leq \pi/2$, the step and node structure are smeared out when increasing the radius of the obstacles. Eventually, in the limit $L_x \gg L_y$, lateral transport is appreciable only for $\phi > \pi/2-\phi^*$. Under such drive conditions, the curve $\mu_{\perp}(\phi)$ boils down to a brad peak, which is seemingly independent of $L_x$. Correspondingly, the longitudinal mobility, $\mu_{\parallel}$, is suppressed and the longitudinal diffusivity grows orders of magnitude larger than $D_0$.

\section{Summary} \label{conclusion}

Forced transport across a 2D array was investigated for a simplified model where a single overdamped Brownian particle of negligible size is suspended in an unmovable interstitial fluid at fixed temperature. The particle is free to diffuse in the connected space delimited by circular reflecting obstacles of finite radius, arranged in a rectangular lattice. The particle is only subject to thermal fluctuations and a homogeneous constant driving force.

Brownian particle transport in the stationary regime was investigated by analyzing the dependence of the particle mobility and diffusivity on the external drive (magnitude and orientation) and the array geometry (obstacle radius and lattice constants). Novel transport properties, including negative differential mobility, excess diffusion peaks, and unconventional asymptotic behaviors, which went unnoticed in earlier reports, have been detected by means of extensive numerical simulations. Such properties have been explained in terms of two distinct depinning mechanisms: (i) trapping by a single obstacle, and (ii) correlated collisions against obstacle rows at an angle with the external force. The corresponding length-scales have been identified to be, respectively, the obstacle radius (independent of the array geometry) and the effective obstacle spacing (resulting from the combination of array geometry and drive orientation).

All other effects, including asymmetry of the obstacles and of their spatial arrangement, particle-particle and particle-obstacle interactions, hydrodynamic corrections, chaotic and inertial dynamical terms, particle size and shape, have been ignored in this report. The generalization of our analysis to incorporate inertia and hydrodynamic corrections is the next step of this research project.

Despite the various approximations and simplifications discussed above, the effects investigated here are robust enough to challenge experimenters investigating the diffusion of extended objects \cite{MacDonald,Huang,Han,Moshchalkov,Drazer1,trap arrays,Renzoni} or even point-like charge carriers \cite{Long} in 2D arrays. Moreover, the interplay of different diffusion length scales has been recently invoked also to explain certain features of the long-time self-diffusion of spherical tracer particles in periodic porous nanostructures \cite{ACS}.

\section*{Acknowledgements}

This work was partly supported by the European Commission under grant No. 256959 (NANOPOWER), the Volkswagen foundation projects I/83902 and I/83903, the German excellence cluster ``Nanosystems Initiative Munich'' (NIM) and the Augsburg center for Innovative Technology (ACIT) of the University of Augsburg. FN was partially supported by LPS, NSA, ARO, NSF grant No. 0726909, JSPS-RFBR contract No. 09-02-92114, Grant-in-Aid for Scientific Research (S), MEXT Kakenhi on Quantum Cybernetics, and the JSPS via its FIRST program. We thanks the RIKEN RICC for providing computing resources.

\end{document}